\documentclass[prl,twocolumn,a4paper,superscriptaddress]{revtex4}

\usepackage{amssymb}
\usepackage{bm}
\usepackage{epsfig}
\usepackage{graphicx}
\usepackage{amsmath}
\usepackage{color}
\newcommand{\blue}{} 
\newcommand{\red}{} 
\usepackage{epstopdf}

\def\sq23{23$\times \sqrt{3}$}

\begin{document}

\preprint{APS/123-QED}

\title{Quadratic X-ray Magneto-Optical Effect in Near-normal Incidence Reflection}

\author{S. Valencia}\email[Corresponding author: ] {sergio.valencia@gmx.de} \affiliation{Helmholtz-Zentrum-Berlin, BESSY, Albert-Einstein-Str.\ 15, D-12489 Berlin, Germany} 
\author{A. Kleibert}  \affiliation{Swiss Light Source, Paul Scherrer Institut, CH-5232 Villigen, Switzerland}
\author{A. Gaupp} \affiliation{Helmholtz-Zentrum-Berlin, BESSY, Albert-Einstein-Str.\ 15, D-12489 Berlin, Germany}
\author{J. Rusz} \affiliation{Department of Physics and Materials Science, Uppsala University, Box 530, S-751 21 Uppsala, Sweden}
\author{D. Legut} \affiliation{Department of Physics and Materials Science, Uppsala University, Box 530, S-751 21 Uppsala, Sweden}
\author{J. Bansmann} \affiliation{{\blue Institute of Surface Chemistry and Catalysis}, Universit\"at Ulm, D-89069 Ulm, Germany}
\author{W. Gudat} \affiliation{Helmholtz-Zentrum-Berlin, BESSY, Albert-Einstein-Str.\ 15, D-12489 Berlin, Germany}
\author{P.M. Oppeneer} \affiliation{Department of Physics and Materials Science, Uppsala University, Box 530, S-751 21 Uppsala, Sweden}

\date{\today}

\begin{abstract}

We have observed a quadratic x-ray magneto-optical effect in near-normal incidence reflection at the $M$ edges of iron. 
The effect appears as the magnetically induced rotation of $\sim$0.1$^\circ$ of the polarization plane of linearly polarized x-ray radiation upon reflection.
{\blue A comparison of the measured rotation spectrum with results from x-ray magnetic linear dichroism data demonstrates that this is the first observation of the Sch\"afer-Hubert effect in the x-ray regime. \textit{Ab initio} density-functional theory calculations reveal that hybridization effects of the $3p$ core states necessarely need to be considered when interpreting experimental data. The discovered magneto-x-ray effect holds promise for future ultrafast and element-selective studies of  ferromagnetic as well as  antiferromagnetic materials.}
\end{abstract}

\maketitle


X-ray magneto-optical spectroscopy techniques are widespread, sensitive methods for element-selective characterization of magnetic systems~\cite{StohrSiegmann}. In particular the great sensitivity of resonant magnetic scattering methods has been demonstrated in many experiments exciting the $L_{3,2}$ edges {\blue ($2p \rightarrow 3d$ transitions)} of $3d$ transition metals (TM) in magnetic nanostructures with system sizes down to the atomic scale~\cite{Gambardella2003}. These experiments have recently been extended from static investigations towards magnetization dynamics~\cite{Stamm2007}. While the temporal structure of synchrotron radiation restricted the time resolution to nanoseconds in the past, studies on the ultrafast magnetization dynamics have become nowadays feasible using femtosecond x-ray pulses provided, e.g., by novel femtoslicing facilities at third generation synchrotron radiation sources such as the ALS (Berkeley, USA), BESSY (Berlin, Germany), and the SLS (Villigen, Switzerland)~\cite{schoenlein00}. Much higher photon flux and thus improved experimental sensitivity in magneto-optical experiments will become available with the advent of soft x-ray free electron lasers (FEL). However, the existing FEL facility FLASH (Hamburg, Germany) is designed to provide photon energies of up to 200~eV, that is, the $L_{3,2}$ edges  of $3d$ TM (in the range of 650 to 950~eV) are currently not accessible.\newline
\indent An alternative is provided by resonant {\blue 3\textit{p}~$\rightarrow$~3\textit{d} transitions, i.e.,} the $M$ edges (at 50 to 65~eV), where the observable magneto-optical effects may possess almost the same order of magnitude when compared to the $L_{3,2}$ edges (see, e.g., {\blue Refs.~\cite{hecker2005,ValenciaNJP}}). Moreover, the importance of the $M$ edges for the investigation of TM compounds might reach soon beyond large scale facilities. Berlasso {\it et al.} \cite{Berlasso} have recently demonstrated the feasibility of performing ultrafast, {\blue\textit{table-top}} experiments at the $M$ edges of TM through the higher order harmonic generation (HHG) of fs laser pulses. Consequently, ultrafast, element-selective magneto-optical techniques {\blue exciting the $3p$ core level electrons} can become accessible to most laboratories.
Despite of these promising properties the $M$ edges are rarely investigated so far and their capabilities for the above described experiments have not yet been explored. {\blue To fully profit from current FEL capacities and future HHG possibilities for element-specific static and time-resolved magnetization studies it is then necessary to further explore magneto-optical techniques in this promising energy range.}  

In this Letter we report the discovery of a novel quadratic x-ray magneto-optical effect at the $M$ edges of TM occuring upon reflection of linearly polarized radiation in near-normal incidence. By comparison with {\blue additional x-ray magnetic linear dichroism (XMLD) measurements and} \textit{ab initio} calculations we show that the reported effect is the x-ray analogon to a similar observation made by Sch{\"a}fer and Hubert in the nineties using visible light \cite{Schafervoigt} which subsequently proved to be a valuable tool for the visualization of magnetic domains \cite{Schaferbook}.

The Sch{\"a}fer-Hubert effect results from the symmetry-breaking that occurs due to the {\blue preferred} magnetization axis in a magnetically ordered material.
As a consequence, the indices of refraction are different for linearly polarized light propagating with electric polarization $\mbox{\boldmath$E$}$  parallel to $\mbox{\boldmath$M$}$ ($n_{||}$) and perpendicular to $\mbox{\boldmath$M$}$ ($n_{\perp}$), respectively. Light traversing the material with $\mbox{\boldmath$E$}$ and $\mbox{\boldmath$M$}$ at an angle of $45^{\circ}$ contains equal components $E_{||}$ and $E_{\perp}$. 
In near-normal incidence reflection the magnetic modification embodied in $n_{||}$ and $n_{\perp}$ leads to the magnetic Sch{\"a}fer-Hubert rotation of the polarization plane upon reflection,  which, using Fresnel theory, can be expressed as
\begin{equation}
\theta_\mathrm{SH}\approx  {\rm Re} \left[ \frac{(n_{||}-n_{\perp})n_0}{n_{||}n_{\perp} -n_0^2} \right]
\approx {\rm Re} \left[ \frac{(\epsilon_{||} -\epsilon_{\perp})n_0}{(n^2 -n_0^2)n}\right],
\label{eq1}
\end{equation}
where $n = (n_{||}+n_{\perp})/2$,  $\epsilon_{||}$, $\epsilon_{\perp}$ are the permittivities for $\mbox{\boldmath$E$}||\mbox{\boldmath$M$}$, $\mbox{\boldmath$E$}\perp \mbox{\boldmath$M$}$, respectively, {\red and $n_0$ is the refractive index of the cap layer.}
The dominating quantity for the effect is $\Delta$= $ \epsilon_{||}- \epsilon_{\perp}$ which also is essential to the XMLD and the x-ray Voigt effect that are both observable in transmission \cite{XMLDOppeneer03}. 
Earlier investigations proved that $ \epsilon_{||}- \epsilon_{\perp}$ is to lowest order proportional to $\left<M^2\right>$ \cite{MertinsPRL}. Therefore the Sch{\"a}fer-Hubert effect can be observed for ferromagnetic (FM) as well as antiferromagnetic (AFM) materials. \newline
\begin{figure}[t]
\centering
\includegraphics[width=1.0\columnwidth]{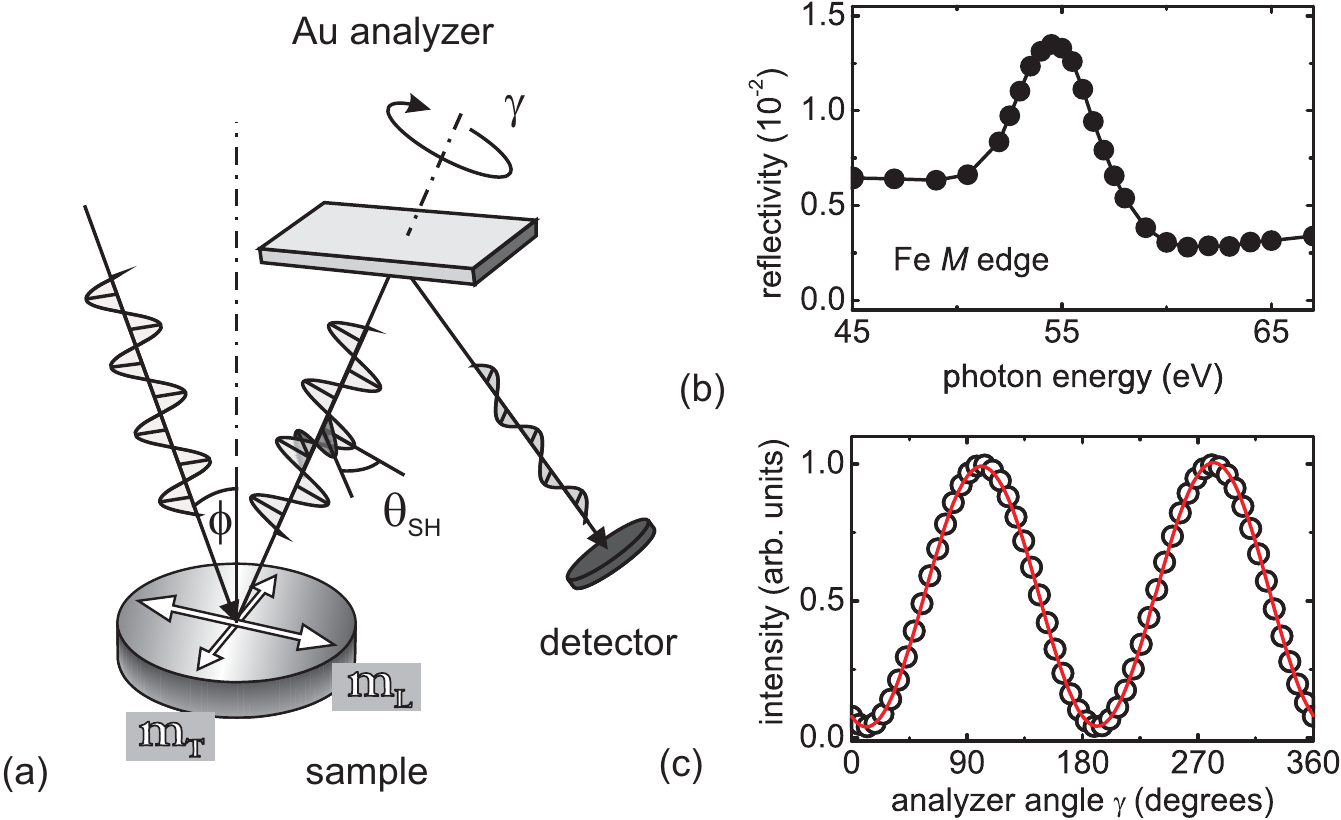}
\caption{(Color online) (a) Experimental setup for the detection of the Sch\"afer-Hubert effect.
(b) Reflectivity in the vicinity of the iron $M$ edges at near perpendicular incidence. (c) Normalized detector signal when rotating the analyzer by an angle of $\gamma$.
}
\label{exper}
\end{figure}
\indent The experiments were performed at the U125/PGM beamline of the synchrotron radiation source BESSY \cite{U1252}. The spectral resolution $E/\Delta E$ was set to 5000. The degree of linear polarization of the incident light was P$_{Lin}$ = 0.99 \cite{U1252}. The BESSY polarimeter chamber \cite{polarimeter} was used for acquisition of the data. In order to guarantee a solely magnetic origin for $\Delta$, i.e. $\Delta$= $ \epsilon_{||}- \epsilon_{\perp}\propto \left<M^2\right>$ [{\blue cf.} Eq.~(\ref{eq1})] the sample must necessarily be cubic or amorphous \cite{kunes03}. The investigated sample was a magnetron-sputtered 50 nm thick Fe layer deposited on a 100~nm thick Si$_3$N$_4$ membrane. A 3~nm Al cap layer was deposited to prevent oxidation. The in-coming radiation was set at an angle of incidence of $\phi=10^{\circ}$ with respect to the {\blue sample} normal as shown in Fig.~\ref{exper}(a). Two magnetic coils allowed magnetic saturation of {\blue$\mbox{\boldmath$M$}$} along two orthogonal directions in the sample plane. The polarization state of the reflected light was analyzed using a rotatable gold mirror analyzer and measuring its reflected intensity at the detector being a GaAs:P photodiode. 

In the present experiments a differential detection scheme was employed in order to detect the 
rotation of the polarization plane of the radiation upon reflection, i.e. the Sch\"afer-Hubert effect. We profit from the general property of quadratic MO effects which in the present case can be expressed by $\theta(\alpha) = \theta_\mathrm{SH}\sin2\alpha$, where $\alpha$ is the angle between the incident polarization and the magnetization as stated above. The measured rotation is thus maximized when setting the magnetization at angle of $\pm\pi/4$ with respect to the polarization. The respective experimental orientations are denoted by $m_\mathrm{T}$ and $m_\mathrm{L}$, respectively, in Fig.~\ref{exper}(a). Subtracting the corresponding rotation angles of the polarization plane then yields $2\theta_\mathrm{SH}$.
Note that due to the deviation of $10^\circ$ from normal incidence an additional small linear MO Kerr rotation may be present in the data. It is eliminated by reversing the magnetization at each orientation and averaging the measurements.

The near-normal reflectivity of the sample in the vicinity of the $M$ edges is depicted in Fig.~\ref{exper}(b). It shows a peak at about 55~eV that accounts for the resonant Fe $3p \rightarrow 3d$ transitions. Contrary to the well studied $L_{3,2}$ edges the core level spin-orbit interaction is much smaller and does not allow to resolve the $M_{2}$ and $M_{3}$ edges separately. {\red The reflectivity is of the order of $10^{-2}$. At the $L_{3,2}$ edges the reflectivity in near normal incidence would be several orders of magnitude smaller, and hence, be below current detector capabilities. Therefore the only feasible detection of an element-selective Sch{\"a}fer-Hubert effect are the $M$ edges}. 
Fig.~\ref{exper}(c) shows the intensity at the detector when rotating the analyzer by an angle of $\gamma$ from 0 to $2\pi$ (open circles) {\blue at an off-resonant energy}. Any {\blue additional} magnetically-induced rotation $\theta$ of the polarization plane causes a 
shift of this curve according to $I(\gamma)=R_0\cdot \left[ 1+P\cdot \cos 2\left(\gamma +\theta\right) \right]$, where $R_0$ denotes the product of the reflectivity of the sample with that of the Au analyzer and \textit{P} is the product of the polarizing power of the Au layer and the degree of linear polarization of the reflected radiation. Here we can 
set $P=1$ \cite{MertinsPRL}. Fitting the above equation to the data [cf.\ the red line in Fig.~\ref{exper}(c)] we obtain $\theta$ for each photon energy and {\blue magnetization}. The Sch\"afer-Hubert rotation is finally given by $\theta_\mathrm{SH}=[\theta(m_\mathrm{T})-\theta(m_\mathrm{L})]/2$.

\begin{figure}[b]
	\centering
		\includegraphics[width=0.43\textwidth]{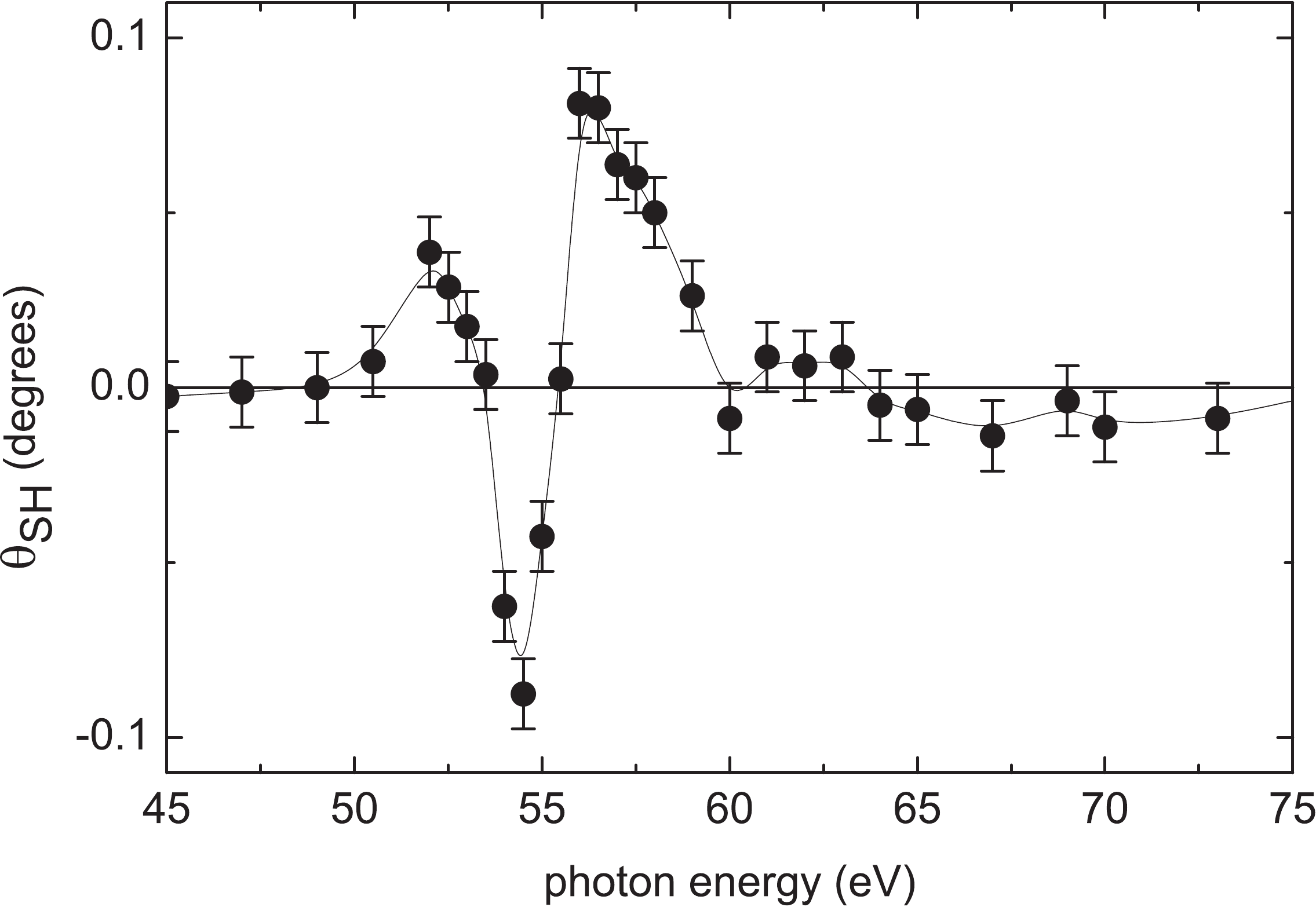}
	\caption{{\blue Photon e}nergy dependent Sch\"afer-Hubert rotation of the polarization plane of linearly polarized x-rays in the vicinity of the Fe \textit{M} edges.}
	\label{fig:SHrotation}
\end{figure}

\begin{figure}[b]
	\centering
		\includegraphics[width=0.43\textwidth]{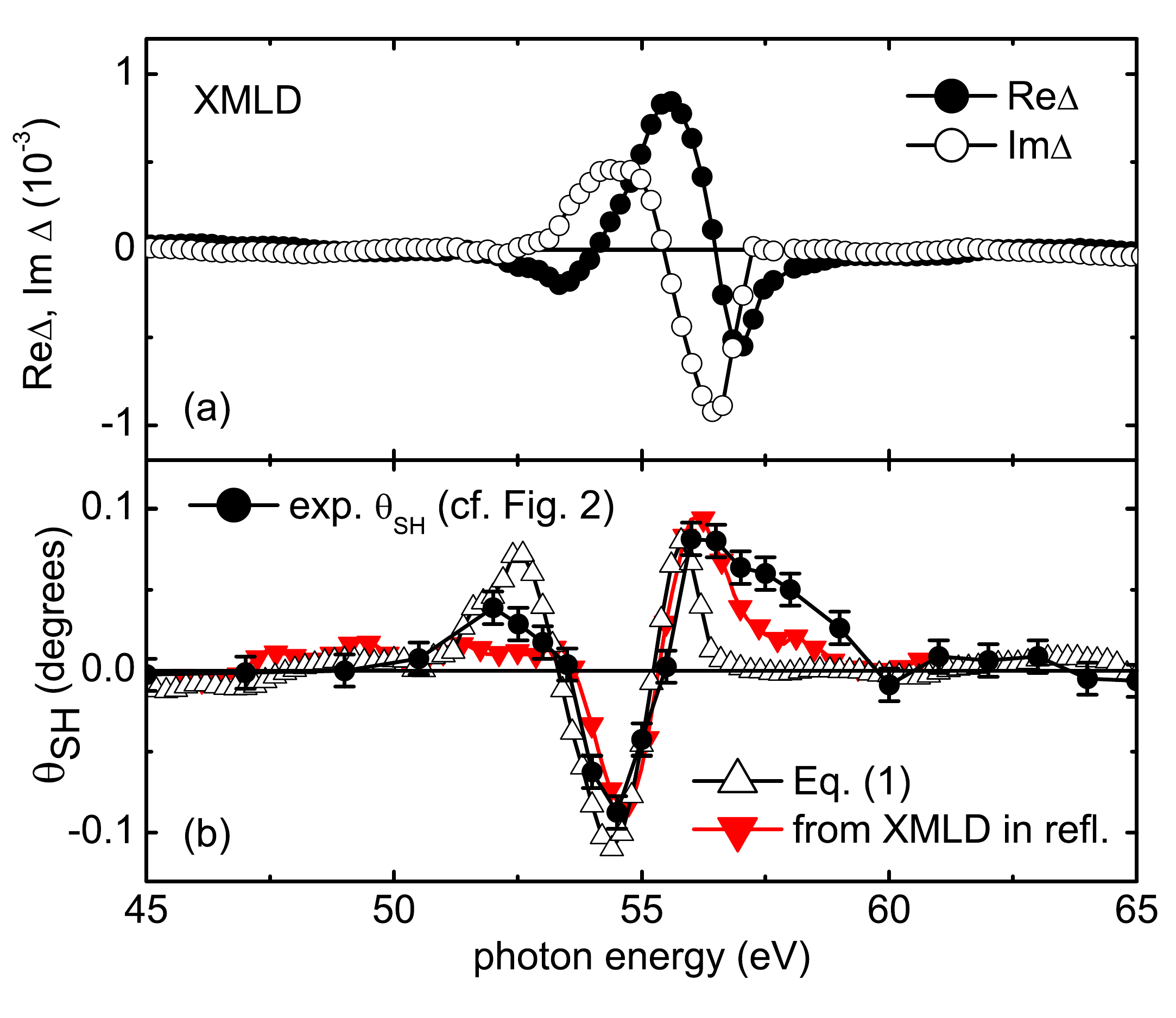}
	\caption{(Color online) (a) Imaginary and real part of the x-ray magnetic linear dichroism measured at the Fe $M$ edges. (b) Calculated Sch\"afer-Hubert rotation $\theta_\mathrm{SH}$ based on (i) the XMLD transmission data of Fig.\ \ref{fig:comparisonXMLD}(a) (open triangles) and (ii) the {\blue XMLD asymmetry measured in reflection} (red triangles). Both data are compared to the experimentally determined $\theta_\mathrm{SH}$ values (solid circles, cf.\ Fig.~\ref{fig:SHrotation}).}
	\label{fig:comparisonXMLD}
\end{figure}

The {\blue resulting} Sch{\"a}fer-Hubert rotation spectrum $\theta_\mathrm{SH}$ is given in Fig.~\ref{fig:SHrotation}. It 
shows a resonant behavior with a twofold sign reversal close to the $M$ edges {\blue and maximum values of about $\pm0.1^\circ$. For comparison we have measured the corresponding XMLD effect in transmission geometry~\cite{MertinsPRL}.}
{\blue Figure~\ref{fig:comparisonXMLD}(a) shows $\rm{Im}\,\Delta$ (open circles) being directly deduced from the transmission data 
and $\rm{Re}\,\Delta$ (solid circles) obtained from a Kramers-Kronig transformation. Using} 
Eq.~(\ref{eq1}) and {\blue the experimental} $\Delta$ values together with reported {\blue data}~\cite{Berlasso} for the permittivity $\epsilon$, 
{\blue yields} the theoretically expected Sch{\"a}fer-Hubert rotation. 
{\blue As depicted in Fig.~\ref{fig:comparisonXMLD}(b) the calculated (upper white triangles) and measured $\theta_\mathrm{SH}$ (solid circles) spectra agree nicely}. {\blue It is worth to mention that} the XMLD data also allow us to deduce a maximum x-ray Voigt rotation in transmission of 8$^\circ$$/\mu$m at the $M$ edges, which is remarkably similar to that measured at the Co $L$ edges {\blue (7.5$^\circ /\mu$m)~\cite{MertinsPRL}.} {\red This is surprising, since, in the conventional understanding it is the larger spin-orbit splitting of the core $j_{3/2}$ and $j_{1/2}$ levels,  being nearly a factor ten larger at the $L$ edges than at the $M$ edges, that is believed to be responsible for the large $L$ edge magneto-x-ray effects. As we will show below through {\it ab initio} calculations, the microscopic mechanism leading to the XMLD at the $M$ edges is actually quite different from that at the $L$ edges.}
The {\red size of the rotation} demonstrates that quadratic magneto-optical effects at the $M$ edges can serve as an equal alternative to respective experiments at the $L$ edges. \newline
\indent {\blue In addition, our experiments confirm a recently predicted relation between the Sch{\"a}fer-Hubert rotation and the XMLD effect occuring in reflection~\cite{XMLDOppeneer03}. The corresponding asymmetry is given by} $A_{\rm{R}} = (R_{\perp} - R_{||}) / (R_{\perp} + R_{||})$ with $R_{\perp}$ and $R_{||}$ being the reflectivity for the magnetization perpendicular or parallel to the polarization plane. At near normal incidence it has been shown that $A_{\rm{R}}=2 \theta_{\rm SH}$~\cite{oppeneer03}. A {\blue respective} $\theta_\mathrm{SH}$ spectr{\red um} computed from the experimentally determined {\blue $A_{\rm{R}}$ data} is given in Fig.~\ref{fig:comparisonXMLD}(b) (red triangles). The agreement with the measured $\theta_\mathrm{SH}$ rotation is again excellent as it reproduces the experimentally measured rotation spectrum both in shape and magnitude. \newline
\begin{figure}[t]
	\centering
		\includegraphics[width=0.43\textwidth]{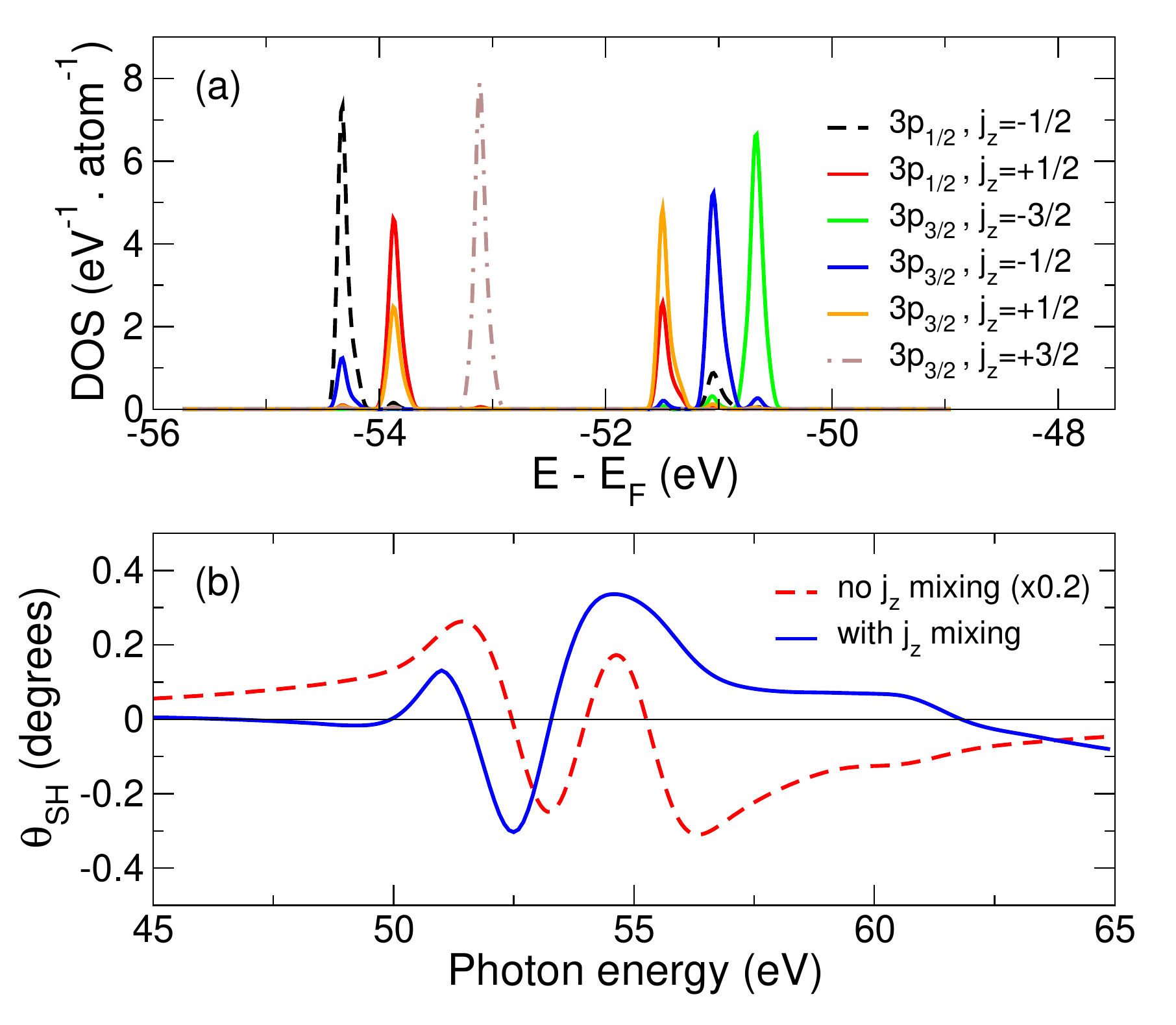}	
	\caption{(Color online) (a) $j, j_z$ resolved density of states (DOS)  of the $3p$ core levels of Fe. (b) Calculated Sch\"afer-Hubert rotation spectrum at the Fe $M$ edge, with $j, j_z $ hybridization included ({\red full} curve) or without (dashed curve).
	}\label{fig:calculations}
\end{figure} 
{\blue {\it Ab initio} density functional theory calculations have been performed using a full-potential linearized augmented plane wave (FLAPW) method in the \textsc{WIEN2k} implementation~\cite{wien2k}. We may note that} a particular difficulty for the theoretical description of the $3p$ semi-core states is related to the relative sizes of the exchange splitting and spin-orbit splitting of the $3p$ levels. Whereas at the $L$ edges the exchange splitting of the $2p$ states is quite small and, consequently, can be treated as a perturbation to the spin-orbit split $j_{3/2}$ and $j_{1/2}$ levels, this can no longer be done for the $3p$ states. In our relativistic calculations {\blue exchange and spin-orbit splitting were therefore included on an equal footing. Also, 
a considerable hybridization of the $j_{1/2}$ and $j_{3/2}$ states can be expected at the $3p$ level. To allow for this, the $3p$ states of iron have been treated as valence states in our calculations}. The combined effect of the exchange and spin-orbit interaction as well as of hybridization on the $3p$ states is illustrated in Fig.~\ref{fig:calculations}{\blue(a)}, where we show the computed $3p$ density of states (DOS). Clearly, the $3p$ states are {\blue not anymore} separate $j_{1/2}$ and $j_{3/2}$ levels, but are mixtures of all $|jj_z\rangle$ components, a situation which is markedly different from that of the $2p$ levels.
Our relativistically calculated energies of the $3p$ levels are in good agreement with a previous calculation \cite{bansmann}, which, however, did not consider the hybridization of the $|jj_z\rangle$ components.\newline
\indent To obtain the Sch{\"a}fer-Hubert rotation spectrum we first calculated the complex dielectric tensor of bcc Fe and subsequently applied the four-vector Yeh formalism \cite{yeh} to obtain $\theta_{\rm SH}$.  
The {\blue theoretically derived} $\theta_{\rm SH}$ spectrum shown in Fig.~\ref{fig:calculations}{\blue(b)} agrees well with the experimental one. {\blue Respective simulations show that the smaller magnitude of the experimental data relative to the theory is due to the Al capping layer, which has been neglected in our calculations given in Fig.~\ref{fig:calculations}(b). }
Note that in the experimental procedure of reversing the magnetizations $m_{\rm T}$ and $m_{\rm L}$ we cancel out a constant background signal. {\blue The corresponding background has accordingly been subtracted for the theoretical $\theta_{\rm SH}$ rotation. \newline
\indent To evaluate the influence of $j,j_z$ mixing in the $3p$ states we have computed $\theta_{\rm SH}$ also without including the $j,j_z$ hybridization. As shown {\red in} Fig.~\ref{fig:calculations}(b) this leads to significant deviations from both the experimental as well the calculated data with hybridization. 
{\red Apart from the} different spectral shape the magnitude {\red of $\theta_{\rm SH}$} is about five times larger than the one with $j,j_z$ mixing included. This demonstrates that for proper description and interpretion of x-ray magneto-optical effects at the $M$ edges it is essential to take the hybridization of the $j, j_z$ levels into account.} \newline
\indent {\blue In conclusion, we have detected a novel quadratic magneto-x-ray effect occuring upon reflection of linearly polarized x-rays in near-normal incidence. A comparison with x-ray magnetic linear dichroism data and \textit{ab initio} calculations confirms that this effect is the x-ray analogon of the Sch\"afer-Hubert in the visible light regime. These calculations also show that the hybridization of the $3p$ core level states has to be considered for a proper description of magneto-optical effects at the $M$ edges of the $3d$ transition metals. With the recent advances in the development of short wavelength optics~\cite{EUVmicrosc} and the increasing availabity of ultrashort UV and X-ray pulses~\cite{Morlens} the Sch\"afer-Hubert effect offers promising opportunities for ultrafast and element-specific microscopy of ferromagnetic and antiferromagnetic domains.} \newline
\indent {\blue S. V. and A. K. contributed equally to this work.} We gratefully acknowledge financial support by the German Federal Ministry of Education and Research (BMBF) Grant No. 05 KS4HR2/4, the Swedish Research Council (VR), STINT, the European Community's Seventh
Framework Programme (FP7/2007-2013) under grant agreement No.\ 214810, 
{}``FANTOMAS\char`\"{}, and the Swedish National Infrastructure for Computing (SNIC). 

\end{document}